\documentclass[12pt]{iopart}
\usepackage{iopams}
\usepackage{graphicx}
\usepackage{amssymb}
\usepackage{appendix}
\usepackage{hyperref}
\usepackage{graphicx}
\usepackage{color}

\newcommand{\be}{\begin{equation}}
\newcommand{\ee}{\end{equation}}
\newcommand{\bea}{\begin{eqnarray}}
\newcommand{\eea}{\end{eqnarray}}

\newcommand{\bef}{\begin{figure}}
\newcommand{\enf}{\end{figure}}

\begin{document}

%\title{New routes for chiral discrimination with high harmonic generation}
%\title{Routes for chiral discrimination with high harmonic generation}
\title{Opportunities for chiral discrimination using high harmonic generation in tailored laser fields}
\author{Olga Smirnova$^{1}$, Yann Mairesse $^{2}$, Serguei Patchkovskii$^{1}$}

%\maketitle

\address{ $^{1}$ Max-Born Institute for Nonlinear Optics and Short Pulse Spectroscopy, Max-Born-Strasse 2A, D-12489 Berlin, Germany}
\address{$^{2}$ CELIA, Universit\'e de Bordeaux - CNRS - CEA, F33405 Talence, France}
%$^{2}$ Department of Physics, Imperial College London, South Kensington Campus, SW7 2AZ London, United Kingdom\\
%$^{3}$ Institute f{\"u}r Physik, Humboldt-Universit{\"a}t zu Berlin, Newtonstrasse 15, 12489 Berlin}

\begin{abstract}
Chiral discrimination with high harmonic generation (cHHG method) has been introduced in the recent work by R. Cireasa et al ({\em Nat. Phys.\/} {\bf 11}, 654 - 658, 2015). In its original implementation, the cHHG method works by detecting high harmonic emission from randomly oriented ensemble of chiral molecules driven by elliptically polarized field, as a function of ellipticity.  Here we discuss  future perspectives in the development of this novel method, the ways of increasing chiral dichroism using tailored laser pulses,  
new detection schemes involving high harmonic phase measurements,
and concentration-independent approaches. 
Using the example of the epoxypropane molecule 
C$_3$H$_6$O (also known as 1,2-propylene oxide),  
we  show theoretically that application of  two-color counter-rotating elliptically polarized laser fields yields an order of magnitude enhancement of chiral dichroism compared  to single color elliptical fields. We also describe how one can introduce a new functionality to cHHG:
concentration-independent measurement of the enatiomeric excess in a mixture of randomly oriented left-handed and right-handed molecules. Finally, for arbitrary configurations of  laser fields, we connect the observables of the cHHG method to the  amplitude and phase of chiral response, providing a basis for reconstructing wide range of chiral dynamics from cHHG measurements, with
femtosecond to sub-femtosecond temporal resolution.  
%Our work opens the way to identifying electron rearrangement times in multielectron systems, detecting the impact of charge dynamics on ionization delays  in time-resolved experiments.
\end{abstract}
%\pacs{42.50.Hz, 32.80.Rm, 33.80.Wz }
\submitto{\jpb}
\maketitle

\section{Introduction}
Ever since their discovery, chiral molecular systems puzzled and inspired researches. The mirror symmetry, characterizing two molecular enatiomers, is possibly the  simplest broken symmetry to think of. The complexity and variety of interactions and occurrences in nature of different enatiomers, flowing from such a small difference in their structure,  is amazing. Chiral nature of living matter poses both fundamental and practical questions, from the origins of homochirality in biomolecules \cite{bonner2000parity} to  detection and manipulation \cite{shapiro2000coherently} of chiral properties. 
%Because a left enantiomer can not be turned into the right one by 
%rotation,  a rotationally isotropic medium that contain  
%enantiomers of a single kind remains chiral. 
The oldest method used to detect chirality in such media is optical rotation: the polarization plane of a linearly polarized light propagating through left and right chiral media rotates in opposite directions 
(see e.g. optical rotation measurement in epoxypropane molecule;  we shall use this molecule to illustrate our concepts here \cite{popyleneoxide-OR-reference}),% see
%e.g. E. Hecht, Optics, 3rd Ed., Addison-Wesley, 1998.) }

Microscopically, optical rotation is governed by the interplay of dipole transitions in a chiral medium caused by electric and magnetic fields of a light wave. While the electric transition dipoles are identical in the two enantiomers in their respective molecular frames, the magnetic dipoles are pointing in opposite directions leading to opposite dynamical electronic 
response in the two enantiomers. 
The same mechanism underlies other linear chiroptical techniques which use circularly polarized light to detect differences in absorption (circular dichroism),  Raman scattering,  or  circular fluorescence \cite{doi:10.1021/ja00437a003} between the two enantiomers. It is useful to note that while chirality is
a topological property of molecular structure, chiral response to light is caused 
by chiral electronic dynamics, which thus effectively 
'measures' this structural property.

Weak magnetic effects lead to weak chiroptical response. 
For example, circular dichroism is often three to six orders of magnitude smaller than light absorbance at the same wavelength.  Weakness of chiral response poses challenges for time-resolved measurements
of chiral dynamics.

%%HHG and alignment, non-randomness of chiral random ensemble

One  way of increasing chiral response is to employ techniques that do not rely on weak  interaction with the magnetic field component of the electromagnetic wave, such as e.g.  
photoelectron  spectroscopy \cite{garcia2013nc}, microwave detection \cite{patterson2013enantiomer}, or Coulomb explosion imaging  \cite{Pitzer06092013}. Two of these techniques,  one-photon \cite{Ritchie76a,Bowering2001,Garcia2003} %[doi:10.1021/jz4014129]
 and multiphoton  \cite{lux2012circular,Lux2015,Powis2013_multi}  
photoelectron circular dichroism (PECD)  are pertinent to the 
approaches to chiral discrimination  involving high 
harmonic generation (cHHG method \cite{chiral}). In particular, 
one-photon ionization is the inverse of  photorecombination, 
which is the key step in high harmonic generation. 

The PECD detection requires resolving the direction of the 
final electron  momentum: the 
photoelectron angular distribution obtained from 
randomly oriented chiral  molecules shows asymmetry with respect 
to  the direction of light propagation. This asymmetry originates 
from the electron interaction with the chiral potential of the core \cite{Ritchie76a,powis2008photoelectron}, 
i.e. is the dynamical consequence of the 
chiral structure. The effect is most 
prominent for low energy electrons \cite{Powis2000_theor}, 
since interaction with the chiral core potential is maximized 
in this case. 
The chiral dichroism signal achievable with PECD, up to some 10\%, sets  the 'golden standard' for  the cHHG method \cite{chiral}.
%We show that using bicircular counter-rotating laser fields \cite{}, cHHG method can match these high values of chiral dichroism. 

With the advent of ultrafast light sources, there has been 
an increasing interest in developing ultrafast chiroptical spectroscopic techniques to investigate chiral response in the time domain. 
In the condensed phase, several breakthroughs have been achieved \cite{fischer05,fidler2014dynamic} using non-linear chiroptical spectroscopy \cite{fischer05,fidler2014dynamic,Mukamel2006,Choi2008}. In the gas phase, where isolated compounds can be studied, such measurements are a lot 
more challenging due to low density of samples, so that no 
ultrafast measurement at $\sim 10^1$ fsec time scale has 
yet been reported. High sensitivity of PECD may open a path towards such measurements, provided that ultrashort circularly polarized laser \cite{lux2012circular,Lux2015,Powis2013_multi} or XUV pulses \cite{ferre2014table,fleischer2014spin} are used. 
 
 Importantly, cHHG naturally offers sub-femtosecond temporal resolution and allows one to detect multielectron chiral dynamics on their natural attosecond timescale. The chiral response can be monitored with $\Delta t\sim$ 0.1 fsec resolution \cite{chiral},  orders of magnitude better than achieved so far (see e.g. \cite{rhee2009femtosecond}). Moreover, already 
the first experiments \cite{chiral} demonstrated 
very high enantio-sensitivity of the cHHG approach, leading to
about 2-3\% chiral signal for nearly linearly
polarized laser fields, with ellipticity $\epsilon\sim 1\%$. 
Thus, cHHG may represent an 
exciting alternative to PECD for ultrafast studies in the gas phase. Here we discuss the opportunities for extending the first cHHG measurements \cite{chiral} to increase their sensitivity and versatility, striving to turn cHHG into 
a standard ultrafast chiroptical detection technique.
 
High-harmonics generation (HHG) can be understood as a sequence of three steps \cite{corkum93,schafer1993prl}:  ionization in a strong laser field, 
laser-induced acceleration of the liberated electron, and its recombination with the parent ion, all within the same laser cycle. 
High harmonic generation spectroscopy can be viewed as a pump-probe technique. Ionization acts as a pump, simultaneously launching the continuum electron and the correlated multielectron dynamics in the cation.
Recombination acts as a probe of such dynamics \cite{Baker06a,Smirnova09c,haessler2010attosecond}. The pump-probe delay is controlled by light oscillation with attosecond precision; the ionization and emission times can be accurately characterised \cite{mairesse2003attosecond,ohio,Shafir12a}. 
Effectively, the emitted light records a movie of the recombining system, 
with each harmonic representing a single frame 
\cite{Baker06a,Smirnova09c}. 

The ability to record multielectron dynamics on attosecond time-scale \cite{Smirnova09c,haessler2010attosecond} is one of the most exciting 
features of high harmonic spectroscopy. This multielectron dynamics can be visualised as the motion of a hole created 
in a molecule upon ionization.  The shape, location, and momentum of the hole during recombination 
are mapped onto the properties of the emitted light - its amplitudes, phases, and polarizations \cite{Smirnova09c,Smirnova09a,Smirnova09-1,mairesse10}. This link is crucial for cHHG.

Indeed, one of the effective mechanisms of chiral sensitivity of 
HHG is directly linked  to attosecond laser-driven dynamics of the hole 
between ionization and recombination. Equivalently, one can think 
about laser-driven multi-electron dynamics in the molecular ion, 
i.e. light-induced transitions between its electronic
states. 
Chiral sensitivity arises due to  the interference of two types of transitions between the electronic states of the cation \cite{chiral}: 
the electric dipole transitions driven by the minor component of 
the electric field, and the magnetic dipole transitions driven by 
the  major component of the magnetic field. 
This inteplay of the magnetic dipole and electric
dipole transitions occurs on the sub-cycle time-scale. The  
enantio-sensitive sub-cycle dynamics reflects the
molecular structure, which dictates the strength 
of the strong-field driven chiral response. 
Note that this dynamics
occurs with the nuclei still located at their positions in the neutral molecule: there
is very little time for the nuclei to move during the 
fraction of the laser cycle between 
ionization and recombination. 

Optimal regime for the interference of the 
magentic dipole transition  driven by 
the  major component of the magnetic field 
and the electric dipole transition
driven by the minor component of 
the electric field requires balance between the 
magnitudes of the respective interactions, $|\mu H_0|\sim|\epsilon d E_0|$. Here $E_0=H_0$ are the amplitudes of the electric and magnetic fields of the laser pulse, $\mu$ is the 
matrix element of the magnetic dipole transition, $d$ is
the matrix element of the electric dipole transition,
$\epsilon$ is the ellipticity of the laser field, $\epsilon E_0$ is
the amplitude of the minor component of the electric
field. Since $|\mu|\sim10^{-2}|d|$, the balance is achieved 
for $|\epsilon|\sim 1\%$, i.e. for nearly linear fields. 
Since $ \vec{\mu}$ changes direction by $180^{o}$ between the 
two enantiomers of the chiral molecule, constructive interference requires opposite values of helicity for right and left enantiomers,
giving rise to chiral dichroism.
 Chiral dichroism at a few percent level for 
such almost linear fields observed in Ref. \cite{chiral} 
demonstrates high potential for chiral 
discrimination of this new technique. 
However, as with any new technique, its first implementation  
may  not be optimal.

The  observed cHHG signal  \cite{chiral} 
maximises for very low ellipticities $\epsilon\sim 1\%$.  
On the one hand, the chiral response typically increases with increasing  ellipticity, likely maximizing for circular  rather than nearly linear fields.  On the other hand,
the  harmonic signal drops quickly with increasing ellipticity. These conflicting requirements seem to preclude exploration of what could be a more favourable regime.
Resolving this conflict should enhance sensitivity and 
flexibility of the cHHG technique.

Tailoring driving fields to maximize the chiral response and optimize the cHHG is a clear direction forward. Specifically, one should aim to
(i) maximise effects of the chiral hole dynamics in the molecular ion between ionization and recombination,
without sacrificing high harmonic generation efficiency;
(ii) explore application of other tools of high harmonic 
spectroscopy to chiral discrimination, 
complementing  measurements of the harmonic spectra 
with  measurements of the harmonic phase \cite{Smirnova09c,haessler2010attosecond,mairesse2003attosecond,
bertrand2013linked}. 
Below, we show how both aims can be achieved using tailored multicolor fields. 

The use of 
multicolor, shaped fields for cHHG  should allow one to achieve 
two important goals.

First, additional fields can couple ionic states populated during ionization to the ionic 
states not directly populated during the ionization step, but characterized by strong magnetic 
dipole transitions, for example due to the proximity of their respective orbitals to a chiral center.
This would make the cHHG method truly  general, applicable to many chiral molecules, even if the
ionization step does not directly populate electronic 
states with strong magneto-dipole transitions.

Second, circular light has long been used to detect chiral response 
because such light has high chirality. 
However, the  high harmonic yield maximises for linear fields 
and quickly drops with increasing $\epsilon$. Application of 
chiral multi-color fields resolves this
problem. 
The family of counter-rotating two-color fields,
introduced in \cite{milovsevic2000attosecond}, can combine circular or  elliptic polarization with strong high harmonic response \cite{fleischer2014spin}. 

Here we  consider the application of  elliptic 
counter-rotating two-color laser fields,
\begin{equation}
E_1=E_0(\epsilon_1\cos(\omega t)\widehat{e}_x+\sin(\omega t)\widehat{e}_y),\\
E_2=E_0(\epsilon_2\cos(2\omega t)\widehat{e}_x-\sin(2\omega t)\widehat{e}_y),
\label{field}
\end{equation}
to induce, enhance and  manipulate the chiral HHG response 
from randomly oriented 
gas of chiral molecules. 
In Eq.(1) 
 $E_0$ is the field strength, $\epsilon_1$, $\epsilon_2$ are the ellipticities of the fundamental field and its second
harmonic, $\widehat{e}_{x,y}$  are the unit vectors in the
poarization (x,y) plane of the laser field.

Below, we theoretically show  that application of such fields should allow one (i) to improve the sensitivity of the cHHG method to rival the sensitivity of the PECD measurements,
without relying on using the driving light with very small ellipticity, (ii) to increase the range of harmonics for which the chiral signal is strong. 
We also show how phase measurements of the chiral high harmonic signal allow one to  determine the amount of specific enatiomers (enantiomeric excess) in a mixture,
in a way that does not require one to know the medium concentration. 
Finally, we present the theoretical background for the reconstruction of time-dependent chiral response for arbitrary ellipticities and 
configurations of the driving laser fields.

%We will show that these features can be used 
%for both imaging chiral response \cite{chiral} and identifying chiral structures with sensitivity approaching PECD measurements.

%Michanism of chiral sensitivity
% The mechanism of chiral sensitivity is based on the interference of dipole transitions 

%Values for epoxypropane

%The ways of increasing chiral dichroism

\section{Chiral discrimination based on HHG spectral and phase measurements}
\label{ii}
In this section, we apply the approach we have already used in \cite{chiral} to estimate the high harmonic signal vs ellipticity of the field given by Eq. (\ref{field}).
We set the ellipticities of both fields
to be equal, $\epsilon_1=\epsilon_2=\epsilon$. We will consider
an epoxypropane molecule CH$_3$CHCH$_2$O (also known as 1,2-propylene oxide), which has been used  
 in the first cHHG experiments \cite{chiral}.

There are two key cHHG channels  in strong field ionization of epoxypropane. These channels are associated with leaving the ion either in the ground electronic state $X$ upon ionization, or in the first excited state electronic state $A$. In the former case, the returning electron can recombine with the ion in state $X$, in the latter case, with the ion in state $A$. 
Higher excited electronic states are
well separated in energy from these two, and do not 
contribute to ionization in strong mid-infrared laser fields.

Recombination of the continuum electron
with the parent ion in these states gives raise to direct 
HHG channels, $XX$ and $AA$. Here 
the first letter marks the ionization channel and the last marks 
the recombination channel.
Laser-induced transitions between the states $X$ and $A$ of the ion open additional cross-channels \cite{mairesse10}: $XA$ and $AX$.
For example, strong field ionization can leave the ion in the 
state $X$, but laser-induced transitions can excite the ion to the 
state $A$. The returning electron now has to recombine with this state: the hole created in the molecule during ionization is different from the hole before recombination.  This results
in the $XA$ channel in HHG.

Importantly, $XA$ and $AX$ are chiral sensitive since 
they involve the interference of the 
laser-driven electric dipole and  magnetic 
dipole transitions \cite{chiral}.    
Our calculations \cite{chiral} show that ionization and recombination create strongly preferred direction for the $ A$ channel, while 
the ionization-recombination angular dependence for the channel $X$ has no such preferred direction. 
Also, the cross-channel $XA$ associated with  
the ionization to the channel $ X$ followed by recombination to the channel $ A$ is 
suppressed relative to the channel $AX$. Along the direction of maximal ionization 
for the channel $ A$, the strength of 
the channel $XA$ is $3\times 10^{-2}$ compared to the
channel  $AX$.

These considerations allow us to develop a very simple 
model of the overall process \cite{chiral}:
\begin{eqnarray}
  \overline{{\bf d}\left(N\right)}=  \overline{{\bf d}_{ X X}\left(N\right)}
+ \overline{{\bf d}_{ A A}\left(N\right)}
+\overline{{\bf d}_{ X A}\left(N\right)}
+ \overline{{\bf d}_{ A X}\left(N\right)}\simeq 
\nonumber \\
\simeq \overline{{\bf d}_{ X X}\left(N\right)}
 +  \overline{{\bf d}_{ A A}\left(N\right)}
 + \overline{{\bf d}_{ A X}}\left(N,\Theta_{max}\right) =
\nonumber \\
= \overline{{\bf d}_{ A A}\left(N\right)}
\left[[1+R_{ A X}]e^{-iE_{ A X}\tau(N)}+R_{ X X}\right]
\label{sfa:model0}
\end{eqnarray}
The symbol ${\bf d}_{IJ}$ denotes 
N-th harmonic component of the laser-induced 
polarization associated with channel $IJ$, i.e. 
ionization that leaves the ion in the state $I$ and
recombination with the ion in the state $J$. 
The overline denotes orientation-averaged quantities.
The  complex-valued relative 
amplitude of the cross-channel, $R_{AX}$, is given by 
orientation-averaged amplitude of the laser-induced 
transition from the state $A$ to the state $X$. This amplitude takes into account all laser-driven transitions, including the 
transitions driven by the magnetic field component due to 
the strong magnetic dipole which couples these states. 
The cross-channel $ A X$ is maximized in the direction given 
by the solid angle $\Theta_{max}$ that optimizes ionization 
into $A$ and recombination into $X$
in the molecular frame. The contribution of the cross-channel is 
still averaged over all angles in the plane orthogonal to
this direction. 

In the last equality in Eq.(\ref{sfa:model0}), we have written down explicitly the relative phase evolution
between the channels $XX$ and $AA$, with $\tau(N)$ being the time-delay between ionization
and recombination and $E_{ A X}\tau(N)$ the difference in the 
energies between the  ionic states $A$ and $X$.  Relevant molecular properties including transition matrix elements are described in detail in \cite{chiral},  where we have performed calculations for the single-color elliptical field.  Here we use the same input and strategy and  only change the single-color 
elliptical field to the field given by Eq. (1).

According to the calculations of
the ionization and recombination amplitudes \cite{chiral}, the relative 
contribution of the $ XX$ channel, 
$R_{XX}$, is close to unity. We have established \cite{chiral} that in the single color elliptical field, due to the weakness of the chiral sensitive cross-channels ($XA$, $AX$), these channels can only leave their fingerprint in the high harmonic spectrum  near a specific harmonic order, for which the main channels $XX$ and $AA$ interfere destructively. The phase term $E_{ A X}\tau(N)$ in Eq.(\ref{sfa:model0}) controls this interference.

Following the model in Eq.(\ref{sfa:model0}), to estimate chiral dichroism in the two-color elliptically polarized counter-rotating  fields, we need to know the ionization $t_i(N)$ and recombination $t_r(N)$ times for each harmonic order $N$. This allows one to calculate the chiral cross-channel: the amplitude of populating the state $X$ at  
the time $t_r(N)$ after the ion was created in the 
state $A$ at the time  $t_i(N)$. These times are the well-known solutions of the saddle point equations for high harmonic generation  in such two-color fields \cite{milovsevic2000attosecond}. Using  Eq.(\ref{sfa:model0}) we can now evaluate the normalized high harmonic signal vs ellipticity $\epsilon_1=\epsilon_2=\epsilon$ (see Eq.(1)). 

Fig. 1 (left panel) shows the calculated chiral dichroism as a function of $\epsilon=\epsilon_1=\epsilon_2$, for $E_0$=0.0239 a.u. (I=$2\times 10^{13}$W/cm$^2$)  
and the carrier frequency $\omega$=0.0224 a.u. ($\lambda=2035$ nm).
High values of chiral dichroism are observed in a wide range of harmonic numbers, not only in the region of the destructive 
 interference of the $XX$ and $AA$ channels, which is 
around harmonic 43. Stronger chiral dichroism in this region 
(about 80 $ \%$) as compared to results obtained in weakly elliptical fields  (about 3 $\%$) \cite{chiral} is due to the order of magnitude higher ellipticity ($\epsilon\sim 0.14$). 

There are two factors  leading to this increase for relatively small ellipticities. Firstly, in the range $\epsilon\ll1$, chiral dichroism is proportional to ellipticity: increasing ellipticity  14 times will respectively increase the chiral dichroism. Secondly,  
for the field used here, 
the amplitude of the transition between the states $A$ 
and $X$ is almost two times higher for $\epsilon\sim 0.14$
then for the single-color field of the same ellipticity.

For higher ellipticities, the disparity between the laser-induced transitions between left and right enantiomers increases and the destructive interference is no longer the decisive mechanism in revealing chiral effects. 
For example, chiral dichroism at the level of about 
20\% is found for  harmonics  35-51 and ellipticity $\epsilon_1=\epsilon_2=1$. Such high values of chiral dichroism are on par with  values achievable with PECD method.

The specific value of chiral dichroism for large ellipticities  depends on many additional factors. One such factor is the
achiral background.
Achiral contributions to the harmonic signal are associated \cite{chiral}
with the 'diagonal' HHG channels, involving the cation staying in the same electronic state between ionization and
recombination. We found that importance of these 
channels in cHHG can be controlled via the Stark 
shift in the two-color field Eq.(1). 
Specifically, most chiral molecules are polar and 
have permanent dipoles in the cationic states. 
Laser interaction with the permanent dipole induces 
the linear Stark shift. For intense laser fields,
these shifts easily approach the $eV$ scale. Such shifts play important role in laser-induced dynamics in the cation between ionization and recombination. Therefore, they can be used to control cHHG by shaping the laser field in the plane of polarization. 
This is precisely what is achieved by varying
$\epsilon_1=\epsilon_2=\epsilon$ in Eq.(1). 

Indeed, for a given electronic state of the cation $J$ with a permanent dipole $\bf d^J$,  the additional phase accumulated in this state 
due to 
its interaction with the laser electric field is: 
\begin{eqnarray}
\phi_{Stark}^J\propto \int_{t_i}^{t_r} 
dt E_0(\epsilon f_x(t) d^J_x+f_y(t) d^J_y).
\end{eqnarray}
Here $f_{x,y}(t)$ describe the sub-cycle 
temporal structure of the electric field in the $x,y$ polarization plane, $d^J_x$ and 
$d^J_y$ are the corresponding
components of the permanent dipole. 
The amplitude $d_{JJ}$ of the achiral contribution to 
the high harmonic field,  associated with this state, acquires 
the corresponding additional 
phase factor, $d_{JJ}\propto e^{i\phi_{Stark}^J}$. 
In strong mid-IR fields $\phi_{Stark}^J\gg 1$. Thus, the phase term $\exp(i\phi_{Stark}^J)$ will 
affect the outcome of coherent averaging of the harmonic response over all molecular 
orientations, in every channel.

Changing $\epsilon$ in Eq.(3) changes $\exp[i\phi_{Stark}^J]$, which 
affects coherent averaging over molecular orientations. 
Fig. 1 (right panel) 
demonstrates the linear Stark-based ellipticity control of the amplitude of the achiral 
channel $AA$ in epoxypropane, using the same theoretical approach as in Ref.\cite{chiral}. 
Clearly, we can manipulate the amplitude of this 
channel within large dynamic range, 
approaching one order of magnitude in 
amplitude for the highest harmonics. 
The  channel $XX$ in epoxypropane shows similar (but not identical) behaviour. 
%Note, that large ellipticities were not achievable in our previous study \cite{chiral}, hence such effect could not be observed.

\begin{figure}[h]
\centering

\includegraphics[width=.49\textwidth]{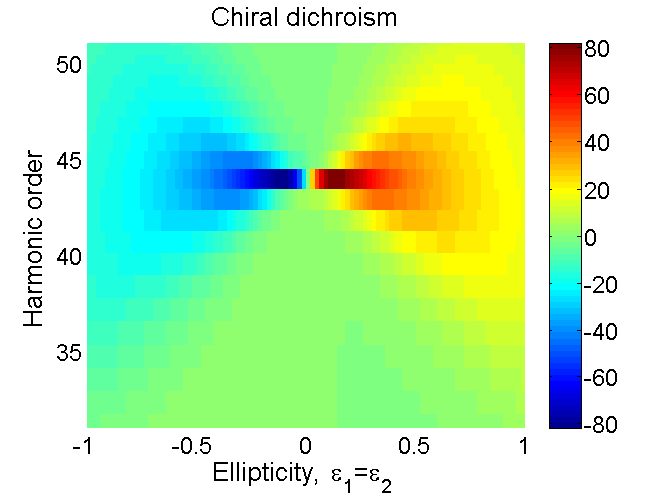}
\includegraphics[width=.49\textwidth]{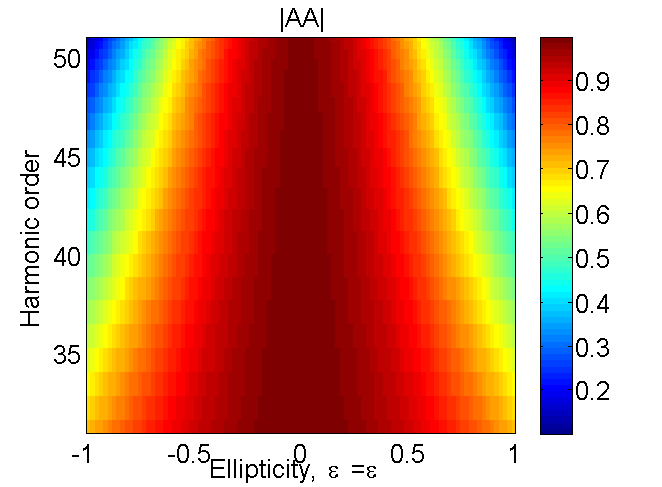}
\includegraphics[width=.49\textwidth]{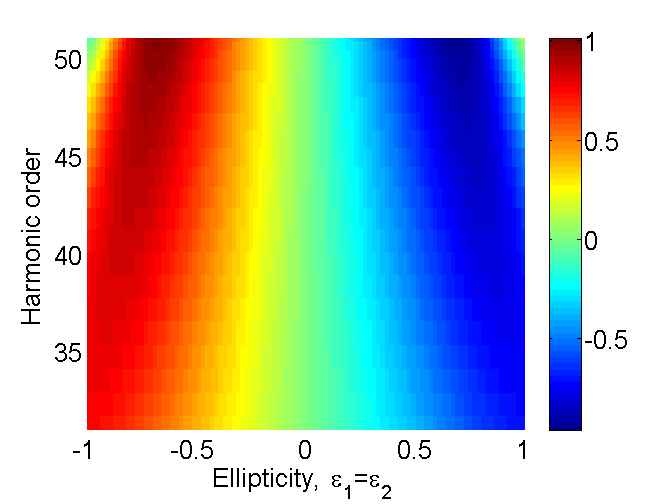}
\sffamily\caption{ HHG in epoxypropane in two-color counter-rotating elliptically polarized fields. Left panel: Estimated chiral dichroism $ Q(N,\epsilon)=2[Y_S(N,\epsilon)-Y_R(N,\epsilon)]/[Y_S(N,\epsilon)+Y_R(N,\epsilon)]$, where $Y_{S,R}$ is the harmonic yield for left (S) and right (R) molecules. 
The color scale is in percentage.
Right panel: The amplitude of $AA$ channel, coherently 
averaged in the direction orthogonal to the direction of maximal ionization of the epoxypropane. Bottom panel: Relative  phase (between the left and right enantiomer)  in radians for the chiral sensitive channel AX in epoxypropane, 
without the contribution of the achiral channels. 
All calculations were done for electric field amplitude $E_0$=0.0239 a.u. ($1.24\times 10^8$V/cm), fundamental frequency
 $\omega$=0.0224 a. u. ($\lambda=2035$ nm), 
 $\epsilon_1=\epsilon_2=\epsilon$.
 %Chiral dichroism $ S(N,\epsilon)=2[Y_S(N,\epsilon)-Y_R(N,\epsilon)]/[Y_S(N,\epsilon)+Y_R(N,\epsilon)]$ for $R_{ X, X}$=0.5. 
}
%\label{chiral_dichroism}
\end{figure}
 
%The second control knob that allows us to manipulate and suppress the chiral background is the laser wavelength. 
%Destructive interference between achiral
%HHG channels helps one to suppress the usually strong achiral background. 
%For two dominant channels, the destructive interference 
%occurs when the 
%relative phase  $\Delta\phi$  accumulated 
%between them is around $\pi$. 
%In the approximation that neglects different photorecombination phases, the phase is $\Delta\phi=\Delta E\tau$, 
%where $\Delta E$ is the energy spacing between the states, $\tau$ is the delay between 
%ionization and recombination. This delay grows with the 
%laser cycle and hence 
%with laser wavelength $\lambda$. Thus, multi-channel interference 
%can be moved around the HHG spectrum by changing $\lambda$. 

\section{Chiral discrimination and chiral dynamic
imaging based on HHG spectral and phase  measurements}
\label{iii}

High harmonic spectroscopy gives access to many observables, including 
harmonic phases \cite{Paul01062001,Smirnova09c,bertrand2013linked}. Chiral sensitivity of the harmonic phase has not been explored so far.
 Fig. 1 (bottom panel) 
shows  the relative phase between 
the chiral-sensitive transition amplitudes in the cation,
for left and right enantiomers. 
This relative phase is substantial, up to one radian, 
well within the accuracy of the experimental phase 
measurements \cite{Smirnova09c}. Thus, cHHG 
can (and should) also aim at using 
phase measurements to detect enantiomers.
In general, achiral channels may
hide this phase. However, here we are
assisted by the suppression of achiral background
shown in Fig.1 (right panel), which indicates that the 
enantio-senitive harmonic phase
could be observable for large ellipticities of the
driving two-color field, where suppression of the
achiral background is strong.  
%%FIGURE1
%\begin{figure}[h]
%%\centering
%\includegraphics[width=.5\textwidth]{phasebetweenAX_channels}
%%\includegraphics[width=.3\textwidth]{background_1}
%%\includegraphics[width=.3\textwidth]{background_0p5}
%\sffamily\caption{\textbf{
%Harmonic phase difference in radians for chiral sensitive channel AX in epoxypropane, $E_0$=0.0239 a.u.,
% $\omega$=0.0224 a. u. 
%}}
%\label{phase}
%\end{figure}
\subsection{Detection of the amount of enantiomers in a mixture.}

The combination of the harmonic phase measurement with the harmonic 
spectral measurement brings the new opportunity:  \textit{concentration-independent} detection of enantiomeric excess in a mixture $M$ of left $S$ and right $R$ chiral molecules (see Fig.2(a)). 
%FIGURE2
\begin{figure}[h]
%\centering
\includegraphics[width=1\textwidth]{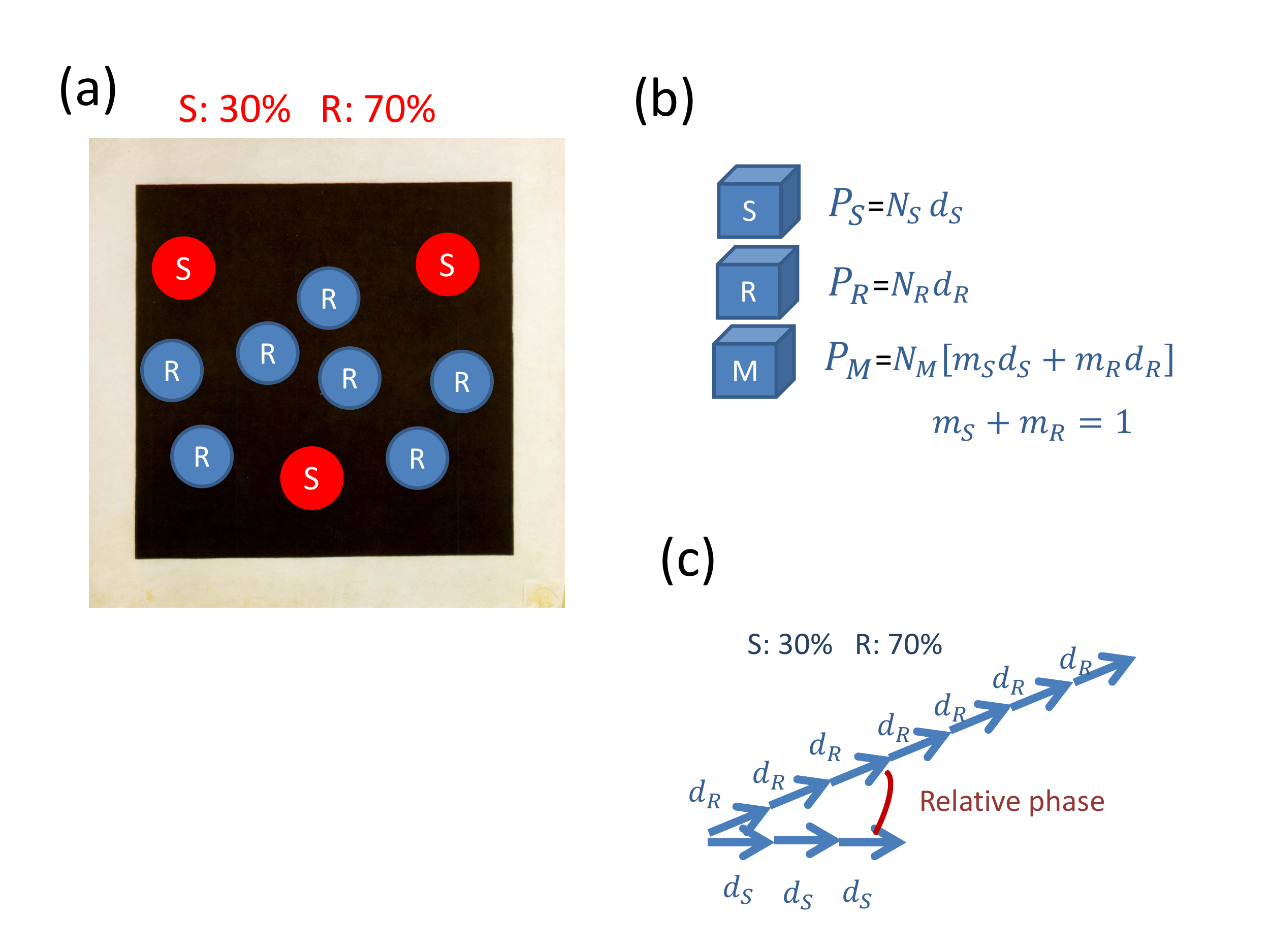}
\sffamily
\caption{
Cartoon illustrating 
importance of the harmonic phase in measuring 
enantiomeric excess using cHHG. 
(a) A mixture of $70\%$ right-handed  and 
$30\%$ left-handed molecules. The goal is to find $\alpha=0.3/0.7$ (b) Schematic representation of the 
harmonic response from three gas cells with pure 
left-handed molecules, pure right-handed molecules, and 
the unknown mixture. The overall number density of the 
mixture is $N_M$, the relative fractions of the 
left-handed and right-handed molecules are $m_S$ and $m_R$, $P_{R,S,M}$ are the polarizations of the media, $d_{R,S,M}$ are the harmonic dipoles. (c) In a mixture, the harmonic dipoles from the 
left-handed and right-handed molecules add coherently, therefore the relative phase between them is crucial for determining $\alpha=m_S/m_R$ and the enatiomeric excess.
}
\label{phase}
\end{figure}
Polarization of the medium is proportional to number density. Therefore, accurate determination of enantiomeric excess in chiroptical spectroscopies requires precise determination of number density, which can be very challenging. In the following we show how one can perform the measurement of enantiomeric excess without any knowledge about number densities of the mixture and of pure samples. Suppose we have a mixture of 
chiral molecules with an overall number density $N_M$ and the 
relative fractions of the left-handed and right-handed molecules $m_S$ and $m_R$. Our goal is to find $\alpha=m_S/m_R$, which
maps into the enantiomeric excess ($e.e.$) as $e.e.=(1-\alpha)/(1+\alpha)\times 25\%$.

The first step is to measure  the
harmonic yield from the unknown mixture, $Y_M(N,\epsilon)$. 
Ideally, this should be done for all $\epsilon$, changing it from 
$\epsilon=0$ to $\epsilon=1$
and then reversing the helicity (denoted as negative $\epsilon$).
At the very least, one should measure the yield for, say, $\epsilon=1$ and $\epsilon=-1$, i.e. perform two spectral measurements.

Next, we use these two measurements
to obtain the 
number density-independent quantity such as  
the elliptic dichroism,
\begin{eqnarray}
ED_{M}(N,\epsilon)=2
\frac{Y_{M}(N,\epsilon)-Y_{M}(N,-\epsilon)}
{Y_{R,S,M}(N,\epsilon)+Y_{R,S,M}(N,-\epsilon)}
\label{Eq:dichroism}
\end{eqnarray} 
or  the ratio of the two harmonic
intensities
\begin{eqnarray}
\beta^2_{M}(N,\epsilon)=\frac{Y_{M}(N,+\epsilon)}
{Y_{M}(N,-\epsilon)}.
\label{Eq:ratio1}
\end{eqnarray}

We can then  compare these measurements 
to the harmonic 
yields  $Y_{R,S}(N,\epsilon)$ from the samples of pure S 
(left-handed) or R (right-handed) molecules, also made 
number density-independent by measuring the corresponding
elliptic dichroisms
\begin{eqnarray}
ED_{R,S}(N,\epsilon)=2
\frac{Y_{R,S}(N,+\epsilon)-Y_{R,S}(N,-\epsilon)}
{Y_{R,S}(N,+\epsilon)+Y_{R,S}(N,-\epsilon)}
\label{Eq:dichroism2}
\end{eqnarray} 
and/or finding the number density-independent ratios
\begin{eqnarray}
\beta^2_{R,S}(N,\epsilon)=\frac{Y_{R,S}(N,+\epsilon)}
{Y_{R,S}(N,-\epsilon)}.
\label{Eq:ratio2}
\end{eqnarray} 
For definitiveness, we assume that $\beta_R>1$, i.e. for
the R molecules the
harmonic yield for $\epsilon$ is higher then for
$-\epsilon$.

Suppose we find that for the pure R-sample and, say, $\epsilon=\pm 1$, the corresponding elliptic dichroism is
$ED_{R}(N,\epsilon=1) =10\%$, and for our mixture we have 
measured $ED_{M}(N,\epsilon=1)=5\%$. 
We also know that for the racemic mixture the dichroism is zero. Can we conclude that our mixture is exactly half-way between pure and racemic? Fig.2 shows why we cannot make such conclusion: the harmonic dipoles from S and R molecules add coherently 
in the mixture. Thus, the total  signal 
also depends on the relative phase $\phi_{SR}$ 
of the harmonic emission between the left-handed and right-handed molecules -- 
the angle between the two harmonic dipoles in Fig.2 (c). 
This phase can be measured using the two-slit harmonic phase measurement technique developed in \cite{Smirnova09c}. 
In this case, the measurement should use far-field 
interference of the 
harmonic light from
two gas cells with enantiomerically pure $S$ and $R$ molecules.
The shift of the intereference fringes relative to
the case of two identical cells 
determines the relative phase $\phi_{SR}$ in a  
concentration-independent way, with  relative
concentrations determining fringe contrast but not fringe
positions. 

Once we have measured $\phi_{SR}$,
we can use it together with the measured $\beta^2_R$ and
$\beta^2_M$ to find $\alpha=m_S/m_R$. Simple algebra 
following addition of the harmonic dipoles shows that
\begin{eqnarray}
\beta_M^2=
\frac{\alpha^2+2\alpha\beta_R\cos\phi_{SR}+\beta_R^2}
{\beta_R^2\alpha^2+2\alpha\beta_R\cos\phi_{SR}+1}
\label{Eq:QuadraticEq}
\end{eqnarray}
Here we have also taken into account that $|d_S(N,-\epsilon)|=
|d_R(N,+\epsilon)|$.

%Let $M=m_1S+m_2R$, where $m_1+m_2=1$ and  $S$ and 
%$R$ denote 
%concentrations of left and right molecules in the mixture. 

%Let us now write the harmonic fields generated in each cell as
%$P_S=N_S d_S$, $P_R=N_R d_R$, and $P_M=N_M (m_1d_S+m_2 d_R)$, 
%where  $d_{S,R}$ denote the microscopic single-molecule response, 
%$N_{S,R}$ are the concentrations of molecules
%in the respective cells, and $m_1$, $m_2$ are the fractions of left and right molecules in the mixture with concentration $N_{M}$, $m_1+m_2=1$. 

%Next, we note that
%\begin{eqnarray}
%\frac{P_{M}(\epsilon)}{P_{M}(\epsilon=0)}=
%\frac{m_1d_{S}(\epsilon)+m_2d_{R}(\epsilon)}{m_1d_{S}(\epsilon=0)+m_2d_{R}(\epsilon=0)}
%=m_1 D_S(\epsilon)+m_2 D_R(\epsilon)
%\label{alpha01}
%\end{eqnarray} 
%since $m_1+m_2=1$ and $d_S(\epsilon=0)=d_R(\epsilon=0)$. $D_{R,S}$ are normalized values of harmonic fields as defined in Eqs. (\ref{alpha00},\ref{alpha01}).

%We now show how  
%$\alpha=m_1/m_2$ can be extracted from the three measurements
%described above. Writing $d_{S,R}=
%|d_{S,R}|\exp(i\phi_{S,R})$, we can extract  $\alpha$ from elliptical dichroism measurements on the mixture and right molecules.
 
%\begin{eqnarray}
%\hspace{-2.5cm}\alpha(N,\epsilon)=\frac{-\beta_R(N,\epsilon)\cos(\phi_{SR}(N,\epsilon))(\beta_M(N,\epsilon)^2-1)}{(\beta_R(N,\epsilon)^2\beta_M(N,\epsilon)^2-1)}+\\\nonumber\hspace{-2.5cm}\frac{\sqrt{\beta_R^2\cos^2(\phi_{SR}(N,\epsilon))(\beta(N,\epsilon)^2-1)^2+(\beta_R(N,\epsilon)^2-\beta_M(N,\epsilon)^2)(\beta_R(N,\epsilon)^2\beta_M(N,\epsilon)^2-1)}}{(\beta_R(N,\epsilon)^2\beta_M(N,\epsilon)^2-1)}.
%\label{alpha5}
%\end{eqnarray}
Varying $\alpha=m_S/m_R$ between $\alpha=0$ (pure
R-sample) and $\alpha=\infty$ (pure S-sample), we see that $\beta_M$ is squeezed between $\beta_R>1$ and 
$\beta_S=1/\beta_R<1$, i.e., $\beta_S=1/\beta_R<\beta_M<\beta_R$.

Next, we re-write the above equation as a 
quadratic equation for $\alpha$ and, taking into
account that  $\beta_S=1/\beta_R<\beta_M<\beta_R$, find
its positive solution:
\begin{eqnarray}
\hspace{-2.5cm}\alpha=\frac{-\beta_R\cos\phi_{SR}(\beta_M^2-1)+\sqrt{\beta_R^2\cos^2\phi_{SR}(\beta_M^2-1)^2+(\beta_R^2-\beta_M^2)(\beta_R^2\beta_M^2-1)}}{(\beta_R^2\beta_M^2-1)}.
\label{alpha5}
\end{eqnarray}
where all quantities depend on the harmonic number, $N$, and
the ellipticity $\epsilon$. 
%Here $\beta_{R,M}=|d_{R,M}^{+}|/|d_{R,M}^{-}|$ is the ratio of harmonic signals  measured for the opposite helicity of the two-color elliptical field. 
%Using the condition $\beta_S=1/\beta_R<\beta_M<\beta_R$ it is easy to show that  $\beta_M^2\beta_R^2>1$ and $\beta_R^2>\beta_M^2$.

%We remind the reader that independence of the measurement on the 
%concentration arises from normalizing harmonic response to its value at 
%$\epsilon=0$, see Eqs.(\ref{alpha00},\ref{alpha01}).
%From now on, all signals will be understood as normalized to their values
%at $\epsilon=0$, thus explicitly removing their dependence on
%concentration of molecules.

Clearly, the main challenge in this approach stems from the
experimental signal to noise ratio, which 
may result in using small quantities with large 
error bars. This challenge can be addressed by taking advantage 
of the  huge redundancy of the two-dimensional 
cHHG measurement, performed for various harmonics N 
as a function of $\epsilon$. Indeed, every observable used in Eq. (\ref{alpha5}) depends on two parameters: $N,\epsilon$. Therefore, the application of
Eq. (\ref{alpha5})  to every harmonic and every ellipticity should yield exactly the same value of $\alpha(N,\epsilon)=m_1/m_2$.

The above scheme 
includes a set of reference measurements 
performed on pure samples. Their outcome  can be used to characterize multiple samples containing unknown mixtures of left and right molecules, under the same experimental conditions. 
The additional measurements, specific to the mixture, are the \textit{HHG intensity} measurements yielding elliptical dichroism
or $\beta_M^2(N,\epsilon)$ for the unknown sample. 

Crucially, identification of $\alpha$ using Eq. (\ref{alpha5}) is concentration-independent: 
one does not need to know the overall number density of molecules in the pure R-sample, and one does not need to know the 
overall number density of molecules in the mixture. 

\subsubsection*{Alternative scheme for detecting 
enantiomeric excess.}

The above measurement scheme for detecting  
enatiomeric excess is not unique. One can employ
%For example, one can use an alternative scheme in which all measurements are done for the same helicity, not for the 
%two opposite values of helicity as described above. 
%The price to pay is 
two additional phase measurements, specific for the mixture, measuring the relative phase between harmonics
emitted by the left-handed molecules and the mixture and the 
relative phase between the harmonics from the right-handed molecules and the mixture. We now consider such a scheme.

Again, we prepare 3 gas cells: with enantiomerically pure left molecules $S$, right $R$ molecules, and the
unknown mixture $M$. Using the two-slit technique,
we now measure three relative phases: (i) 
the relative phase between the harmonics generated from the left-handed and the right-handed samples $\phi_{SR}$, 
(ii) between the left-handed molecules and the 
mixture $\phi_{MS}$, and 
(iii) between the right-handed molecules 
and the mixture, $\phi_{MR}$.
We now define  number density-independent
signals by normalizing the harmonic fields at $\epsilon\neq 0$
to their values at $\epsilon=0$,  
\begin{eqnarray}
\frac{P_{S,R}(\epsilon)}{P_{S,R}(\epsilon=0)}=
\frac{d_{S,R}(\epsilon)}{d_{S,R}(\epsilon=0)}\equiv D_{S,R}(\epsilon)
\label{alpha00}
\end{eqnarray}
Next, we note that
\begin{eqnarray}
\frac{P_{M}(\epsilon)}{P_{M}(\epsilon=0)}=
\frac{m_1d_{S}(\epsilon)+m_2d_{R}(\epsilon)}{m_1d_{S}(\epsilon=0)+m_2d_{R}(\epsilon=0)}
=m_1 D_S(\epsilon)+m_2 D_R(\epsilon)
\label{alpha01}
\end{eqnarray} 
since $m_1+m_2=1$ and $d_S(\epsilon=0)=d_R(\epsilon=0)$.
Here $D_{R,S}$ are the normalized values of harmonic fields as defined in Eqs. (\ref{alpha00},\ref{alpha01}).

We now show how  
$\alpha=m_1/m_2$ can be extracted from the measurements
described above. Writing $D_{S,R}=
|D_{S,R}|\exp(i\phi_{S,R})$, we can explicitly calculate the argument $arg[D_MD_R^{*}]=\phi_{MR}$ in two steps. First,
\begin{eqnarray}
D_MD_R^{*}=\left[ m_1D_S+m_2D_R\right]D^{*}_R=m_1D_SD^{*}_R+m_2|D_R|^2 .
\label{alpha1}
\end{eqnarray}
Second, for the quantity  $D_SD^{*}_R$ in the above equation we write
\begin{eqnarray}
D_SD^{*}_R=|D_S||D_R|m_1\left[\cos(\phi_{SR})+i\sin(\phi_{SR})\right]+m_2|D_R|^2,
\label{alpha2}
\end{eqnarray}
Using these two expressions, we finally obtain
\begin{eqnarray}
\tan(\phi_{MR})=\frac{m_1|D_S||D_R|\sin(\phi_{SR})}{m_1|D_S||D_R|\cos(\phi_{SR})+m_2|D_R|^2}.
\label{alpha3}
\end{eqnarray}
Analogously, we obtain for $\phi_{MS}$:
\begin{eqnarray}
\tan(\phi_{MS})=-\frac{m_2|D_S||D_R|\sin(\phi_{SR})}{m_2|D_S||D_R|\cos(\phi_{SR})+m_1|D_S|^2}.
\label{alpha4}
\end{eqnarray}
Introducing the ratio of the spectral amplitudes of the harmonic signals from left and right molecules $\beta(N,\epsilon)=|D_S(N,\epsilon)|/|D_R(N,\epsilon)|$ and $\gamma(N,\epsilon)=\frac{|\tan(\phi_{MR}(N,\epsilon))|}{|\tan(\phi_{MS}(N,\epsilon))|}$, which can be obtained from the phase measurements, we can express $\alpha=m_1/m_2$ from Eqs. (\ref{alpha3},\ref{alpha4}): 
\begin{eqnarray}
\hspace{-2.5cm}\alpha(N,\epsilon)=\frac{\cos(\phi_{SR}(N,\epsilon))(\gamma(N,\epsilon)-1)+\sqrt{\cos^2(\phi_{SR}(N,\epsilon))(1-\gamma(N,\epsilon))^2+4\gamma(N,\epsilon))}}{2\beta(N,\epsilon)}.
\label{alpha05}
\end{eqnarray}
%Experimentally, $\beta$ is measured by normalizing the harmonic response
%to its value for $\epsilon=0$, 
%Note that the parameter $\beta(N,\epsilon)$ representing the ratio
%of the harmonic signals for  $R$ and $S$ enantiomers in the mixture and assuming equal concentration of 
%can be re-written in equivalent, normalized form:
%\begin{eqnarray}
%\beta(N,\epsilon)=\frac{|D_S(N,\epsilon)|}{|D_S(N,\epsilon=0)|} 
%\frac{|D_R(N,\epsilon=0)|}{|D_R(N,\epsilon)|} ,
%\label{alpha5a}
%\end{eqnarray}
%since formally $|D_S(N,\epsilon=0)|=|D_R(N,\epsilon=0)|$. 

Again, we note that every observable 
in Eq. (\ref{alpha05}) depends on $N$ and $\epsilon$. Therefore, the application of
Eq. (\ref{alpha05})  to every harmonic and every ellipticity should yield exactly the same value of $\alpha(N,\epsilon)=m_1/m_2$, thus reducing the signal to noise ratio. 

We remind the reader that in this scheme, the 
independence of the measurement on 
concentration (number density) arises from normalizing the harmonic response 
to its value at 
$\epsilon=0$, see Eqs.(\ref{alpha00},\ref{alpha01}).
From now on, all signals will be understood as normalized to their values
at $\epsilon=0$, thus explicitly removing their dependence on
molecular concentration (number density).

Note that in the second approach we
perform redundant spectral measurements by measuring
harmonic response for both S- and R-molecules while
varying ellipticity in the full range, $-1<\epsilon<1$, increasing the redundancy of the first scheme by factor two.
%, fundamentally, one needs the 
%following input to determine the enantiomeric excess: the phase
%between the left-handed and right-handed molecules, the ratio of
%the harmonic intensities for left and right molecules, for a
%given ellipticity (each harmonic response has to be normalized to
%zero ellipticity to remove concentration-dependence), and $\alpha
%$. To find these five unknowns one needs to perform at least five 
%measurements. In our first scheme we use minimal number of 
%measurements. In the second scheme, we use six measurements, 
%however the three phases used in this schemes are formally not 
%independent of each other - only two phases out of three are 
%independent.}

\subsection{Time-resolving sub-femtosecond chiral response in full range of ellipticities}
\label{iv}
%In our work \cite{chiral} we have established two mechanisms of chiral sensitivity, related to chiral dynamics in the cation between ionization and recombination and chiral dynamics of the returelectron.
As established in Ref.\cite{chiral}, the main  difference between the cHHG signals coming from the two enantiomers is due to the small magnetic-dipole response in the cation. Thus, the multi-electron wavepacket in 
the cation acquires small correction, which is linear in $\mu=|\mathbf{\overrightarrow{\mu}}|$ (a straightforward result of the first order perturbation theory):
%\begin{equation}
%|\Psi_{\rm R,S}^{(D)}(t)\rangle\rightarrow |\Psi_0^{(D)}(t)\rangle+
% |\Delta\Psi_{\rm R,S}^{(D)}(t)\rangle=|\Psi_0^{(D)}(t)\rangle\pm
%\mu|\delta \Psi_{\rm R,S}^{(D)}(t)\rangle.
% \end{equation}
% 
 \begin{equation}
|\Psi_{\rm R,S}(t)\rangle=|\Psi_0(t)\rangle\pm
\mu|\delta \Psi_{\rm ch}(t)\rangle.
 \end{equation}
Here $|\Psi_0(t)\rangle$ is the time-dependent wavefunction 
describing the dynamics in the cation in the 
dipole approximation, $\pm
\mu|\delta \Psi_{\rm ch}(t)\rangle$ is the chiral 
correction of the time-dependent response, opposite for S and R molecules.
This small correction to the multi-electron dynamics in the cation yields chiral correction to the harmonic 
field 
 \begin{eqnarray} 
D_{\rm R,S}(N,\epsilon)=
\big[
D_{\rm R,S}^{(0)}(N,\epsilon)\pm\mu \delta D_{\rm ch}(N,\epsilon) 
\big].
\label{Eq:DysonHHG22}
\end{eqnarray}
Here we have used the number density-independent signals, which can be obtained
 by normalizing the harmonic fields at $\epsilon\neq 0$
to their values for $\epsilon=0$, 
\begin{eqnarray}
\frac{P_{S,R}(\epsilon)}{P_{S,R}(\epsilon=0)}=
\frac{d_{S,R}(\epsilon)}{d_{S,R}(\epsilon=0)}\equiv D_{S,R}(\epsilon)
%\label{alpha00}
\end{eqnarray}
Here $D_{\rm R,S}^{(0)}(N,\epsilon)$ is the normalized harmonic field in the dipole approximation. 

The first-order in $\mu$ expression is general. 
Its only assumption is the linear response of the cation to 
the weak effect of the magnetic field. 
The dipole $D_{\rm R,S}^{(0)}(N,\epsilon)$  is not
chiral sensitive (once averaged over randomly oriented ensemble), 
$D^{(0)}_{\rm S}(N,\epsilon)=D^{(0)}_{\rm R}(N,\epsilon)=D^{(0)}(N,\epsilon)$.
The chiral-sensitive component is  $\mu \delta D_{\rm ch}(N,\epsilon)$.
The harmonic yield for left and right enantiomers is then:
 \begin{equation}
Y_{\rm R,S}(N,\epsilon)=|D^{(0)}(N)|^2
\left[
1\pm 2\mu \frac{|\delta D_{\rm ch}(N)|}{|D^{(0)}(N)|}\cos(\Phi_{ch}(N))
\right].
\label{Eq:Y}
 \end{equation}
Using Eq.(\ref{Eq:Y}), one obtains for the  chiral dichroism 
$Q(N,\epsilon)=2 \frac{Y_S(N,\epsilon)-Y_R(N,\epsilon)}{Y_S(N,\epsilon)+Y_R(N,\epsilon)}$ (the  difference between the harmonic yields $Y_{S,R}(N,\epsilon)$ for left and right molecules divided by the signal from the racemic mixture):
%\begin{equation}
%Q(N,\epsilon)=\frac{2}{1} \frac{Y_S(N,\epsilon)-Y_R(N,\epsilon)}{Y_S(N,\epsilon)+Y_R(N,\epsilon)},
%\end{equation} 
%yields
% \begin{equation}
%S(N,\epsilon)\simeq2\frac{\epsilon_0(N)}{\sigma^2},
% \end{equation}
% and
 \begin{equation}
Q(N,\epsilon)=4 \mu \frac{|\delta D_{\rm ch}(N,\epsilon)|}
{|D^{(0)}(N,\epsilon)|}\cos(\Phi_{ch}(N,\epsilon)).
\label{Eq:Q}
 \end{equation}
Here $\Phi_{\rm ch}$ is the relative phase between the chiral and achiral contributions to the emission amplitude. 
Thus, the standard measure of the chiral signal,
when applied to high harmonic yield, is proportional to 
$\mu|\delta D_{\rm ch}(N,\epsilon)|\cos(\Phi_{ch}(N,\epsilon))$ -- the chiral part of HHG response.
%he electron recombination amplitude with the 
%chiral-sensitive component of the multi-electron wavepacket in
%the cation.

An exciting goal  is to separately reconstruct 
the  \textbf{amplitude} $\mu |\delta D_{\rm ch}(N,\epsilon)|$ 
and the relative \textbf{phase} $\cos(\Phi_{ch}(N,\epsilon))$ of 
the chiral response.  
This can be done by augmenting 
the measurement of $Q(N,\epsilon)$
with two more measurements: (i) of the relative
harmonic phase $\phi_{SR}(N,\epsilon)$ between the 
$S$ and $R$ molecules, and (ii) of the harmonic amplitude
in the racemic mixture $|D^{(0)}(N,\epsilon)|$.

Eq.(\ref{Eq:DysonHHG22}) allows us to explicitly calculate  $\tan(\phi_{SR})$:
\begin{eqnarray} 
\big[
D^{(0)}-\mu \delta D_{\rm ch} 
\big]\big[
D^{(0)}+\mu \delta D_{\rm ch} 
\big]^{*}\approx|D^{(0)}|^2+\mu\delta D_{\rm ch}^{*}D^{(0)}-\mu\delta D_{\rm ch}D^{(0)*},
\label{Eq:DysonHHG2}
\end{eqnarray}
where we have neglected the second order terms in $\mu$. Thus,
\begin{eqnarray} 
\tan(\phi_{SR})=2\mu\sin(\Phi_{ch})\frac{|\delta D_{\rm ch}|}{|D^{(0)}|}.
\label{Eq:tan}
\end{eqnarray}
Since the sign of $\Phi_{ch}$ becomes important, we specify that $\Phi_{ch}=\arg[\delta D_{\rm ch}^{*}D^{(0)}]$. Adding $\tan(\phi_{SR})^{2}$
and $Q^2(N,\epsilon)/4$ we obtain
\begin{eqnarray}
\tan^{2}(\phi_{SR}(N,\epsilon))+\frac{Q^2(N,\epsilon)}{4}=W^2(N,\epsilon),
 \end{eqnarray}
 where $W(N,\epsilon)$ gives  access to the relative amplitude of the chiral component:
 \begin{eqnarray}
W(N,\epsilon)=2\mu\frac{|\delta D_{\rm ch}(N,\epsilon)|}
{|D^{(0)}(N,\epsilon)|}.
\label{Eq:W}
 \end{eqnarray}
The dependence of $W(N,\epsilon)$ on $|D^{(0)}(N,\epsilon)|$ can be further eliminated by performing  the harmonic measurements in the racemic mixture (rm). Suppose this measurement yields 
the harmonic spectrum $y_{rm}(N,\epsilon)$ (also
normalized to the signal for $\epsilon=0$). 
Then, the amplitude of the chiral component can be obtained as follows:
 \begin{eqnarray}
\mu|\delta D_{\rm ch}(N,\epsilon)|=\frac{W(N,\epsilon)\sqrt{y_{rm}(N,\epsilon)}}{2}.
\label{Eq:Ach}
 \end{eqnarray}
  
On the other hand, combining Eqs.(\ref{Eq:Q}),(\ref{Eq:W}) we obtain the phase between the chiral and achiral components:
\begin{eqnarray}
\cos(\Phi_{ch}(N,\epsilon))=\frac{Q(N,\epsilon)}{2W(N,\epsilon)}.%\hspace{0.2cm}
\label{Eq:phase}
 \end{eqnarray}
 Thus, both the  amplitude and the relative phase of the chiral signal can be obtained from the combination of spectral and amplitude measurements.
 
 The result described by Eqs.(\ref{Eq:W},\ref{Eq:Ach}, \ref{Eq:phase})  is in fact more general then may appear at 
first glance: Eq.(\ref{Eq:DysonHHG22}) may equally apply to chiral dynamics initiated in a neutral molecule and then probed with HHG. This scheme is sketched in Fig.3.
 Such pump-probe scheme could allow one to probe the evolution of chiral component $\mu|\delta D_{\rm ch}(N,\epsilon)|$ in  neutral species during structural changes on a longer, femtosecond time scale. This time scale is determined by the delay between the pump pulse, which excites  chiral dynamics in the neutral molecule, and the  probe pulse which generates the chiral-sensitive high harmonic signal. Up to now, no pump-probe time-resolved measurement of ultrafast chirality has been done in the gas phase due to the absence of sufficiently sensitive technique. cHHG promises to open this direction.    
%Here $\phi_{SR}$ is the harmonic phase difference between the two enatiomers and $Q$ is the measured chiral signal.
%, $Q(N,\epsilon)=\frac{2}{1} \frac{Y_S(N,\epsilon)-Y_R(N,\epsilon)}%{Y_S(N,\epsilon)+Y_R(N,\epsilon)}$.

\begin{figure}[h]
\centering
\includegraphics[width=1\textwidth]{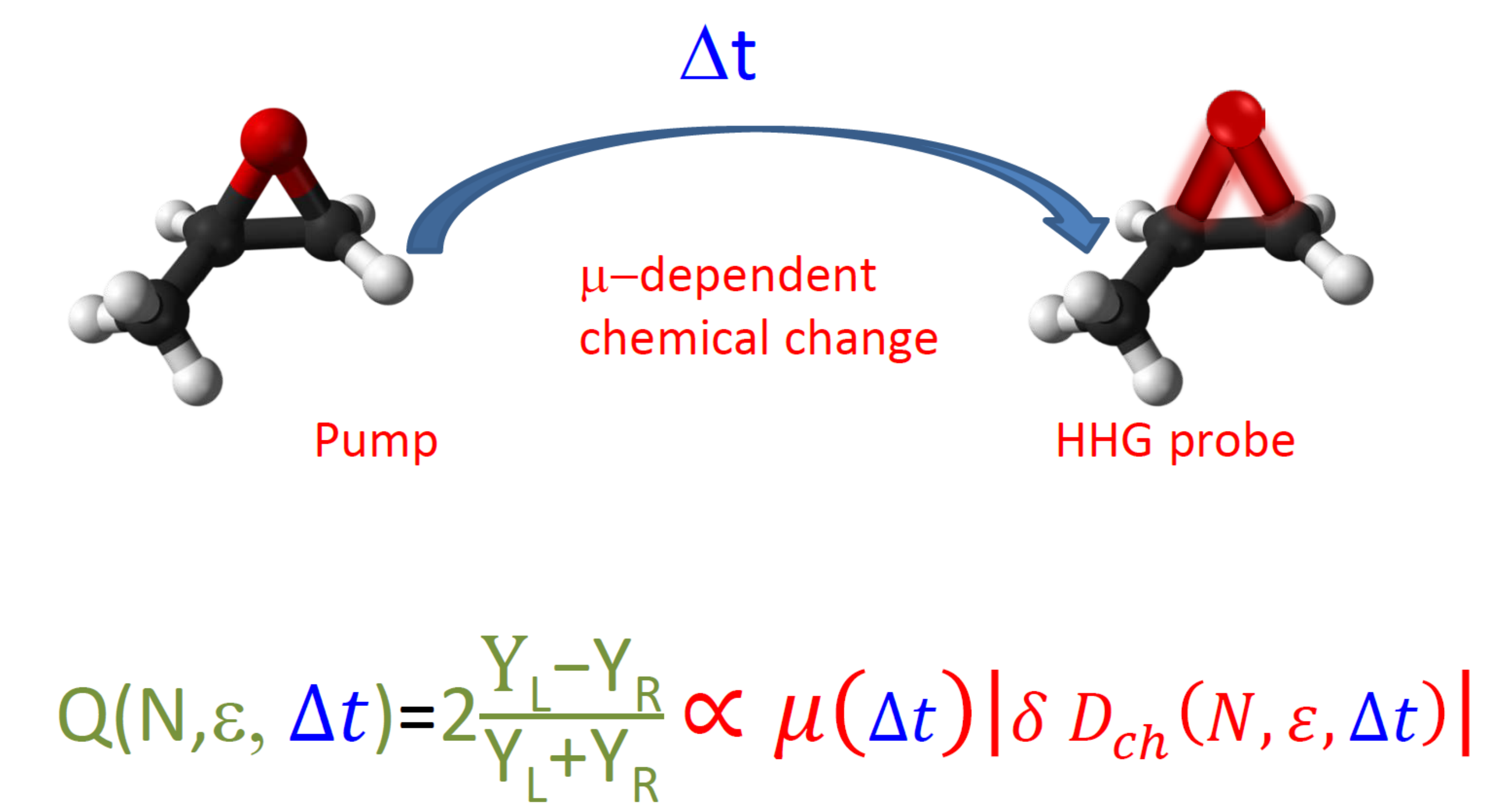}
\sffamily\caption{ A cartoon of pump-HHG probe measurement describing one of the possible schemes of detecting time-dependent  chiral response. The chiral signal in HHG arises from the magnetic dipole. A pump pulse initiates nuclear motion in a chiral molecule. As  a result of this motion, the nuclei rearrange themselves around the chiral center. This leads to changing $\mu$ and therefore changes the high harmonic response induced by the HHG probe.  Thus, HHG allows one 
to follow chiral response from sub fs to fs time scale. 
}
%\label{chiral_dichroism}
\end{figure}

\section{Conclusions}
\label{vii}
Using the example of epoxypropane molecule, we  have 
shown theoretically that cHHG method should strongly  benefit  
from  the application of the 
two-color, counter-rotating elliptically polarized fields. Application of such fields should lead to much higher values of circular dichroism compared to the  original scheme \cite{chiral}, which relied on weakly elliptical fields.

We predict that the combination of the harmonic amplitude and phase measurements should  allow one to directly measure 
enantiomeric excess in a concentration-independent way 
and monitor the evolution of 
the chiral dynamics with sub-femtosecond 
time resolution.
 
We have shown how cHHG method can be used to extract both the  amplitude and  phase of the chiral signal, for arbitrary configurations of  laser fields used to initiate high harmonic process. Indeed, the analysis in Section 3 does not assume any specific field 
configuration.

We predict that Eqs.(\ref{Eq:W},\ref{Eq:Ach},\ref{Eq:phase}) can 
also be used to extract the  amplitude and the phase of the 
chiral signal from the HHG observables in the pump-probe set-up.
In this setup chiral dynamics is first excited in a 
neutral molecule and then probed with  
chiral-sensitive high harmonic signal. From single shot to 
pump-probe, cHHG driven by tailored laser 
fields thus enables imaging of a wide range of chiral dynamics,
with time resolution extended from sub-femtosecond to
many tens of femtoseconds.

\ack
We thank Prof. M. Ivanov for fruitful discussions.
 O. S. and S.P. acknowledge the  support of the DFG grant SM 292/3-1. O.S. gratefully acknowledges the support of the DFG grant 292/5-1.  All authors acknowledge the support of the European COST Action CM1204 XLIC.

\section*{References}

\providecommand{\newblock}{}

%\begin{thebibliography}{10}
% \printbibliography[heading=none]
%\end{thebibliography}{10}

\end{document}